\documentclass[pra,aps,amsmath,amssymb,amsfonts,twocolumn,nofootinbib,longbibliography]{revtex4-1}

\usepackage{amssymb}
\usepackage{}
\usepackage{bm,mathrsfs}

\usepackage{graphicx}
\usepackage{epsfig}
\usepackage{amsmath,bbm}
\usepackage{amsfonts,amssymb}
\usepackage{times}
\usepackage{verbatim}
\usepackage[sort&compress]{natbib}

\usepackage{amsmath}
\usepackage[colorlinks,breaklinks,linkcolor=blue,anchorcolor=blue,citecolor=blue,urlcolor=blue,
dvipdfm
]{hyperref}

\newcommand{\be}{\begin{equation}}
\newcommand{\ee}{\end{equation}}
\newcommand{\bea}{\begin{eqnarray}}
\newcommand{\eea}{\end{eqnarray}}

\def\>{\rangle}
\def\<{\langle}

\def\qed{\leavevmode\unskip\penalty9999 \hbox{}\nobreak\hfill
     \quad\hbox{\leavevmode  \hbox to.77778em{%
               \hfil\vrule   \vbox to.675em%
               {\hrule width.6em\vfil\hrule}\vrule\hfil}}
     \par\vskip3pt}
     
\begin{document}

\newtheorem{theorem}{Theorem}
\newtheorem{lemma}[theorem]{Lemma}
\newtheorem{corollary}[theorem]{Corollary}
\newtheorem{proposition}[theorem]{Proposition}
\newtheorem{definition}[theorem]{Definition}
\newtheorem{example}[theorem]{Example}
\newtheorem{conjecture}[theorem]{Conjecture}
\title{Engineering steady entanglement for trapped ions at finite temperature by dissipation}

\author{Xiao-Qiang Shao}
\email{Corresponding author: shaoxq644@nenu.edu.cn}
\affiliation{Center for Quantum Sciences and School of Physics, Northeast Normal University, Changchun, 130024, People's Republic of China}
\affiliation{Center for Advanced Optoelectronic Functional Materials Research, and Key Laboratory for UV Light-Emitting Materials and Technology
of Ministry of Education, Northeast Normal University, Changchun 130024, China}

\date{\today}

\begin{abstract}
{We propose a dissipative method for preparation of a maximally entangled steady state of two trapped ions in the Lamb-Dicke limit. By addressing the trapped-ion system with a monochromatic standing wave laser pulse of frequency resonant with the ionic transition and a microwave field coupled to the ground-state transitions, we obtain an effective coupling between two particles, which is independent of the phonon-number fluctuations. Meanwhile, the controlled spontaneous emission of trapped ions is implemented via pumping the metastable states upwards to the short-lived ionic states by an auxiliary laser field. Combining the unitary processes with the engineered dissipation, a deterministic Bell state can be produced irrespective of the initial states of systems. Moreover, our result shows that the CHSH inequality can be violated for a wide range of decoherence parameters, even at finite temperature.}
\end{abstract}

\pacs {03.67.Bg, 03.65.Yz, 32.80.Qk, 32.80.Ee}\maketitle \maketitle

\section{introduction}
 In the framework of open quantum system, the idea of quantum dissipation is introduced to deal with the process of irreversible loss of energy observed at the classical level \cite{breuer2002theory}. It is based on the fact that no quantum system is completely isolated from its surroundings, and hence the interaction between quantum system and external environment will lead to dissipation \cite{feynman2000theory}.  Even for a closed quantum system, as long as the total system can be split into two parts: a quantum subsystem of interest and a bath containing an infinite number of degrees of freedom, the energy of the quantum subsystem will unidirectionally flow towards the bath, causing the loss of coherence. Therefore, the effect of quantum dissipation is non-ignorable in  quantum-mechanical system, especially in the filed of quantum information processing, where the high-accuracy quantum operations and quantum states are required.

To fight with decoherence due to the quantum dissipation, the active error-correction approaches and passive
error-prevention schemes are usually employed to protect the quantum information against errors \cite{nature03074,PhysRevLett.100.020502,nature10786,reiter2017dissipative,PhysRevLett.81.2594,PhysRevLett.85.1762,PhysRevLett.107.150401}. For instance, the spin-flip errors of an encoded one-qubit state can be corrected by means of a three-qubit quantum error-correcting code plus syndrome measurements, or the logical qubits are designed to be encoded
into decoherence-free subspaces that are completely
insensitive to specific types of noise. Nevertheless, as we see, the above methods consume a lot of resources and make the quantum system
further susceptible to the environment. Instead of attempting to combat noise, in the seminal work of 2002, Plenio and Huelga utilized an external optical white noise
field to play
a constructive role in the production of entangled light of two leaky optical cavities \cite{PhysRevLett.88.197901}. Since then, the dissipation-based entanglements have been realized in
various light and matter coupling systems, such as cavity QED \cite{PhysRevLett.106.090502,shen2011steady,reiter2012driving,PhysRevA.89.012319,su2014preparation,PhysRevA.90.054302,reiter2016scalable},  optomechanics \cite{PhysRevLett.110.253601,tan2013dissipation,chen2015dissipation}, and superconducting qubits \cite{PhysRevA.88.023849,shankar2013autonomously,leghtas2015confining}.

The first theoretical scheme for preparation of a maximally entangled state of two particles was raised by Kastoryano {\it et al.} in cavity QED \cite{PhysRevLett.106.090502}, where they developed an effective-operator method to illustrate how a singlet state of atoms is prepared as the steady solution of
a dissipative quantum dynamical process, starting from an arbitrary initial state. What's more important, the
cooperativity around $C\approx30$ is good enough to ensure a $\sim90\%$ fidelity with respect to the singlet state, since the cavity decay plays
an integral part in the dynamics. Recently, we loosened the requirement of the cooperativity $C$, in terms of steady entangled state, by combination of quantum Zeno dynamics and atomic spontaneous emission \cite{PhysRevA.95.062339,Li:17}. These proposals are extremely effective as the cavity mode interacting with a zero-temperature bosonic reservior, because the cavity decay can be suppressed by
the Zeno requirement. But for a finite-temperature bath, the heating rate decoherence terms will greatly degrade the entanglement.

The ion trap is a good experimental platform for controlling the quantum degrees of freedom. Similar to the cavity QED, the collective center-of-mass motional mode acts as the quantum bus of the trapped-ion system, which is used
for communication between ions \cite{cirac1995quantum,PhysRevLett.82.1971,PhysRevLett.87.127901,PhysRevA.66.044307,PhysRevLett.90.217901,PhysRevA.68.035801,leibfried2003quantum,schindler2013quantum,monroe2013scaling}. In 2013, Lin {\it et al}. experimentally used continuous,
time-independent fields to engineer a steady entangled state of two qubit ions with fidelity up to
$75\%$, which was then boosted to $89\%$ using stepwise application
of laser fields \cite{nature12801}. In their suggestion, an extra
`coolant' ion has to be introduced for sympathetic cooling of the qubit ions' motion. Lately, Bentley {\it et al}. theoretically enhanced the fidelity of steady entangled ions above $99\%$ by conditioning
the system dynamics to the detection of photons spontaneously
emitted into the environment \cite{PhysRevLett.113.040501}. However, detecting spontaneous emission events is still challenging
using state of the art technology.

In this work, we put forward an alternative scheme for dissipative preparation of a maximally entangled steady state for two trapped ions with neither the sympathetic cooling nor the detecting spontaneous emission. We engineer the spontaneous emission of metastable state of ion  by coupling it to a fast-decaying excited electronic state. This dissipative process, combined with a unitary coherent dynamics robust against the thermal motion of ions, enables us to accomplish a deterministic Bell state of high fidelity, even for a finite-temperature bath.

\begin{figure}
\includegraphics[scale=0.30]{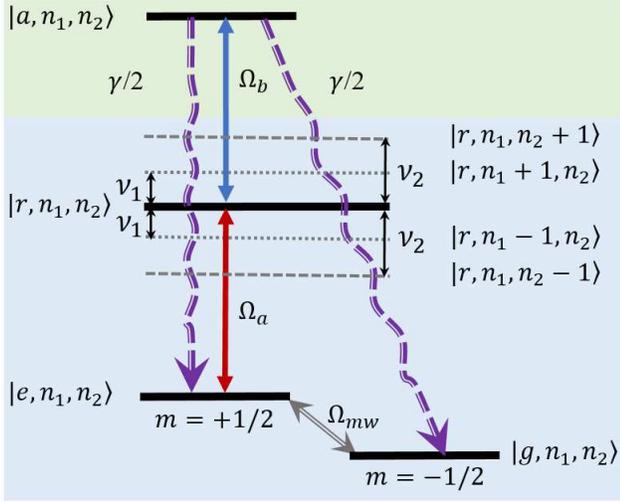}
\caption{\label{fig1}Schematic view of the ionic internal
states. This model incorporates two microwave-field-coupled ground states $|g\rangle$ and $|e\rangle$ (two Zeeman sublevels $m = \mp1/2$), a metastable state $|r\rangle$, and a fast-decaying state $|a\rangle$, and $n_p~(p=1,2)$ represents the quantum number for the relevant
vibrational mode of frequency $\nu_p$. One standing wave laser pulse of Rabi frequency $\Omega_a$ interacts resonantly with the ionic transition $|e\rangle\leftrightarrow|r\rangle$, and another driving field of Rabi frequency $\Omega_b$ pumps the population of ion from $|r\rangle$ to $|a\rangle$. The trap frequency is $\nu$, the lifetime of the metastable state $|r\rangle$ is $\gamma_r^{-1}$ (not shown), and  lifetime of the short-lived state $|a\rangle$ is $\gamma^{-1}$. }
\end{figure}

\section{ Effective model}
The system consists of two four-level ions trapped in a linear trap, whose level structure is characterized in Fig.~\ref{fig1}. $|g\rangle$ and $|e\rangle$ are two ground states corresponding to the Zeeman sublevels $m = \mp1/2$, $|r\rangle$ is a metastable state, and $|a\rangle$ is a temporary (short-lived) level. Without loss of generality, the system we consider is not limited to a particular ion, since the similar configuration can be obtained from simple atomic ions
with a lone outer electron, such as the alkaline-earths
(Be$^+$, Mg$^+$, Ca$^+$, Sr$^+$, and Ba$^+$) \cite{monroe2013scaling}.
 The quantum number and frequency for the $p$th ($p=1,2$)
vibrational modes are marked by $n_p$ and $\nu_p$, respectively.
A monochromatic standing wave of Rabi frequency $\Omega_a$ executes the carrier resonance of the ionic transition $|e,n_1,n_2\rangle\leftrightarrow|r,n_1,n_2\rangle$, which simultaneously induces $\pm\nu_p$ detunings from two sidebands. And another laser pulse of Rabi frequency $\Omega_b$ is in resonance with the carrier transition $|r\rangle\leftrightarrow|a\rangle$ without affecting the motional state. Besides, a microwave field of Rabi frequency $\Omega_{mw}$ is used to couple to the ground-state transition $|g\rangle\leftrightarrow|e\rangle$.

Before discussing the function of the short-lived state $|a\rangle$, we concentrate on the $\Lambda$-type configuration of ion composed by states $|g\rangle$, $|e\rangle$, and $|r\rangle$. In the Schr\"{o}dinger picture, the Hamiltonian of the system reads ($\hbar=1$)
\begin{eqnarray}\label{xxx}
{\hat H_S}&=&\sum_{j=1}^2\sum_{k=r,e,g}\omega_k|k_j\rangle\langle k_j|+ \sum_{p=1}^2\nu_p {\hat a_p}^{\dag}{\hat a_p}\nonumber\\
&&+\sum_{j=1}^2\bigg\{\frac{\Omega_a}{2}\sin\bigg[\sum_{p=1}^2\eta_{jp}({\hat a_p}^{\dag}+{\hat a_p})+\phi\bigg]e^{i\omega_Lt}|e_j\rangle\langle r_j|\nonumber\\
&&+ \frac{\Omega_{mw}}{2}e^{i\omega t}|g_j\rangle\langle e_j|+{\rm H.c.}\bigg\},
\end{eqnarray}
where $\omega_r$, $\omega_e$, and $\omega_g$ are the energies of levels $|r\rangle$, $|e\rangle$, and $|g\rangle$, respectively. ${\hat a_p}^{\dag}$ and ${\hat a_p}$ are the annihilation and creation operators for the $p$th vibrational mode, and $\phi$ is phase of the laser field characterizing the relative position of the
trap center with respect to the node of the laser standing
wave. $\eta_{jp}$ is the Lamb-Dicke parameter incorporating the relative displacement of the $j$th ion in the $p$th mode. $\omega_L=\omega_r-\omega_e$ is the frequency of the standing wave laser pulse and  $\omega=\omega_e-\omega_g$ denotes the frequency of the microwave field. In the Lamb-Dicke regime $\eta_{jp}\sqrt{\overline{n}_p+1}\ll1$, we first consider the case of perfect alignment that the
ions are positioned at the node of the field ($\phi$=0), then the Hamiltonian of the system in the interaction picture can be described by
\begin{eqnarray}\label{full}
{\hat H_I}&=&\sum_{j=1}^2\bigg\{\frac{\Omega_a}{2}|e_j\rangle\langle r_j|\bigg[\sum_{p=1}^2\eta_{jp}({\hat a_p^{\dag}}e^{i\nu_p t}+{\hat a_p}e^{-i\nu_p t})\bigg]\nonumber\\
&&+\frac{\Omega_{mw}}{2}|g_j\rangle\langle e_j|+{\rm H.c.}\bigg\}.
\end{eqnarray}

It is noteworthy that for a two-ions system, the angular frequencies of the center-of-mass mode and the breathing mode are characterized by $\nu_1=\nu$ and $\nu_2=\sqrt{3}\nu$, respectively, where $\nu$ is the trap frequency. In addition, the Lamb-Dicke parameters satisfy $\eta_{11}=\eta_{21}=-3^{1/4}\eta_{12}=3^{1/4}\eta_{22}$ \cite{james1998quantum}. In the situation of $\eta_{jp}\Omega_a/2\ll\nu_p$, there is no energy exchange between the internal state of ions and the external vibrational mode, thus we can obtain a phonon-number-independent Hamiltonian via the second order perturbation method as \cite{PhysRevA.68.035801}
\begin{eqnarray}\label{effective}
{\hat H_{\rm eff}}&=&-\sum_{p=1}^2\frac{{\eta_{1p}\eta_{2p}\Omega_a^2}}{2{\nu_p}}\prod_{j=1}^2(|e_j\rangle\langle r_j|+|r_j\rangle\langle e_j|)\nonumber\\
&&-\sum_{j=1}^2\bigg[\sum_{p=1}^2\frac{{\eta_{jp}^2\Omega_a^2}}{4{\nu_p}}(|e_j\rangle\langle e_j|+|g_j\rangle\langle g_j|)\nonumber\\
&&-\frac{\Omega_{mw}}{2}(|g_j\rangle\langle e_j|+|e_j\rangle\langle g_j|)\bigg].
\end{eqnarray}
Taking the spontaneous emission of the metastable state $|r\rangle$ into account, the Markovian master equation is modeled in the Lindblad form
\begin{equation}
\dot{\hat\rho}_t = i[\hat\rho_t,{\hat H_{\rm eff}}]+\frac{{\gamma_r}}{2}\sum_{j=1}^2 \{\hat {\cal L}[|g_j\rangle\langle r_j|]\hat\rho_t+\hat {\cal L}[|e_j\rangle\langle r_j|]\hat\rho_t\}, \label{Mast}
\end{equation}
where $\gamma_{r}$ denotes the spontaneous decay rate of state $|r\rangle$
and the superoperator $
\hat {\cal L}[\hat c]\hat\rho=\hat c\hat\rho\hat c^{\dag}-\frac{1}{2}\hat c^{\dag}\hat c\hat\rho-\frac{1}{2}\hat\rho\hat c^{\dag}\hat c
$.
The dissipative factor of the phonon has been neglected because the vibration freedoms of ions are decoupled to internal electronic freedom of the ions. Now it is easy to inspect that the antisymmetric Bell state $|S\rangle=(|eg\rangle-|ge\rangle)/\sqrt{2}$ is the unique steady solution of the master equation of Eq.~(\ref{Mast}).

Mathematically speaking, the steady-state population $P_{t\rightarrow\infty}=\langle S|\hat\rho_{t\rightarrow\infty}|S\rangle$ of unity is always attainable in the presence of a non-zero $\gamma_r$, but a smaller $\gamma_r$ must be accompanied by a longer convergence time to make the final population of Bell state sufficiently high. As listed in Table.~\ref{table1}, {we suppose $\eta_{11}=\eta$ and fix the Rabi frequency of the microwave $\Omega_{mw}=-2\lambda$, where $\lambda=-\sum_{p=1}^2\eta_{1p}\eta_{2p}\Omega_a^2/(2\nu_p)=-\eta^2\Omega_a^2/(3\nu)$, and alter the spontaneous decay rates by setting $\gamma_r=\lambda$, $\gamma_r=0.1\lambda$, and $\gamma_r=0.01\lambda$, respectively. Although the final population of all these three cases can increase to unity, a choice of $\gamma_r=\lambda$ is the best since the population of this case has already reached $99.86\%$ at a finite time $\tau=800/\lambda$. In contrast, the selection of $\gamma_r=0.01\lambda$ only has $44.86\%$ population at the same time.} From an experimental point of view, a suitable value of $\gamma_r$ is able to shorten the process of convergence
and hence avoid other unknown decoherence factors as far as possible. For typical trapped-ion systems, the decay rate of the metastable state $|r\rangle$ is on the order of $\sim$Hz \cite{leibfried2003quantum,schindler2013quantum,haffner2008quantum}, which is generally too small to be ignored. In what follows, we will discuss in detail how to manipulate the spontaneous decay rate of the metastable state in a controllable way.
\begin{table}
\centering
\scalebox{1.1}{
\begin{tabular}{p{2cm}p{2cm}p{2cm}l}
\hline\hline
$\Omega_{mw}$ & $\gamma_r$ & $P_{\tau}$& $P_{t\rightarrow\infty}$\\ \hline
$-2\lambda$ &$\lambda$ &0.9986&1.0000 \\
$-2\lambda$ &0.1$\lambda$ &0.7281&$1.0000$\\
$-2\lambda$ &0.01$\lambda$ &0.4486&$1.0000$\\ \hline\hline
\end{tabular}}
\caption{\label{table1} Populations of the antisymmetric Bell state at a finite time $t=\tau=800/\lambda$ and an infinite time $t\rightarrow\infty$ under different spontaneous decay rates of $|r\rangle$.}
\end{table}

\section{Engineered spontaneous emission}
In order to see the mechanism of controlled spontaneous emission clearly, we discard the effective Hamiltonian $\hat{H}_{\rm eff}$ in Eq.~({\ref{Mast}}) and only consider a standing wave laser pulse in the Lamb-Dicke limit ${\hat H^j_{au}}=\Omega_b/2\cos[\sum_{p=1}^2\eta_{jp}({\hat a_p}^{\dag}+{\hat a_p})](|a\rangle\langle r|+|r\rangle\langle a|)\approx\Omega_b/2(|a\rangle\langle r|+|r\rangle\langle a|)$ resonantly driving the single-ion transition $|r\rangle\leftrightarrow|a\rangle$. Supposing the ion decays from
the short-lived state $|a\rangle$ to the ground states $|g\rangle$ and $|e\rangle$ with the
same spontaneous emission rate $\gamma/2$, the master equation can be written as
\begin{equation}
\dot{\hat\rho}_{\rm sin} = i[\hat\rho_{\rm sin},{\hat H^j_{au}}]+\frac{\gamma}{2}\hat {\cal L}[|g\rangle\langle a|]\hat\rho_{\rm sin}+\frac{\gamma}{2}\hat {\cal L}[|e\rangle\langle a|]\hat\rho_{\rm sin}. \label{iMastEqns}
\end{equation}

After expanding the density operator of single ion in the form
$
{\hat\rho}_{\rm sin}(t)=\sum_{k,l}{\hat\rho}_{kl}(t)|k\rangle\langle l|, (k,l=a,r,e,g)
$
and  substituting it into the master equation of Eq.~(\ref{iMastEqns}), we obtain a set of coupled equations for the ionic matrix elements:
\begin{equation}
\dot{\hat\rho}_{aa}=i\frac{\Omega_b}{2}{\hat\rho}_{ar}-i\frac{\Omega_b}{2}{\hat\rho}_{ra}-\gamma{\hat\rho}_{aa},\label{motion1}
\end{equation}
\begin{equation}
\dot{\hat\rho}_{ar}=i\frac{\Omega_b}{2}{\hat\rho}_{aa}-i\frac{\Omega_b}{2}{\hat\rho}_{rr}-\frac{\gamma}{2}{\hat\rho}_{ar},\label{motion4}
\end{equation}
\begin{equation}
\dot{\hat\rho}_{ae}=-i\frac{\Omega_b}{2}{\hat\rho}_{re}-\frac{\gamma}{2}{\hat\rho}_{ae},\label{motion5}
\end{equation}
\begin{equation}
\dot{\hat\rho}_{ag}=-i\frac{\Omega_b}{2}{\hat\rho}_{rg}-\frac{\gamma}{2}{\hat\rho}_{ag},\label{motion6}
\end{equation}
\begin{equation}
\dot{\hat\rho}_{rr}=i\frac{\Omega_b}{2}{\hat\rho}_{ra}-i\frac{\Omega_b}{2}{\hat\rho}_{ar},\label{motion2}
\end{equation}
\begin{equation}
\dot{\hat\rho}_{ee}=\dot{\hat\rho}_{gg}=\frac{\gamma}{2}{\hat\rho}_{aa},\label{motion3}
\end{equation}
\begin{equation}
\dot{\hat\rho}_{re}=-i\frac{\Omega_b}{2}{\hat\rho}_{ae},\label{motion7}
\end{equation}
\begin{equation}
\dot{\hat\rho}_{rg}=-i\frac{\Omega_b}{2}{\hat\rho}_{ag}.\label{motion8}
\end{equation}

For the short-lived state $|a\rangle$ with the decay rate on the order of $\sim$100~MHz, we
can adiabatically eliminate the temporary level in the limiting of $\gamma\gg\Omega_b/2$. Thus
it is reasonable to assume $\dot{\hat\rho}_{ar}=\dot{\hat\rho}_{ae}=\dot{\hat\rho}_{ag}=\dot{\hat\rho}_{aa}=0$ and we have
\begin{eqnarray}
&&{\hat\rho}_{aa}=\frac{\Omega_b^2}{\Omega_b^2+\gamma^2}{\hat\rho}_{rr},\  \
{\hat\rho}_{ae}=-\frac{i\Omega_b}{\gamma}{\hat\rho}_{re},\label{eff3}\\
&&{\hat\rho}_{ar}=-\frac{i\Omega_b\gamma}{\Omega_b^2+\gamma^2}{\hat\rho}_{rr},\ \ {\hat\rho}_{ag}=-\frac{i\Omega_b}{\gamma}{\hat\rho}_{rg}\label{eff4}.
\end{eqnarray}
Then Eqs.~(\ref{motion2})-(\ref{motion8}) can be reformulated as
\begin{eqnarray}\label{eff2}
\dot{\hat\rho}_{rr}&=&-\gamma_{\rm eff}{\hat\rho}_{rr},\  \
\dot{\hat\rho}_{ee}=\dot{\hat\rho}_{gg}=\frac{\gamma_{\rm eff}}{2}{\hat\rho}_{rr},\label{eff2}\\
\dot{\hat\rho}_{re}&=&-\frac{\gamma_{\rm eff}}{2}{\hat\rho}_{re},\  \
\dot{\hat\rho}_{rg}=-\frac{\gamma_{\rm eff}}{2}{\hat\rho}_{rg},\label{eff6}
\end{eqnarray}
where $\gamma_{\rm eff}=\Omega_b^2/\gamma$. If $\Omega_b/2\pi=200$~kHz is adopted, the effective spontaneous decay rate is about $\gamma_{\rm eff}\sim0.4$~kHz, which is much larger than the natural spontaneous emission rate of the metastable state.
\begin{figure}
\scalebox{0.42}{\includegraphics{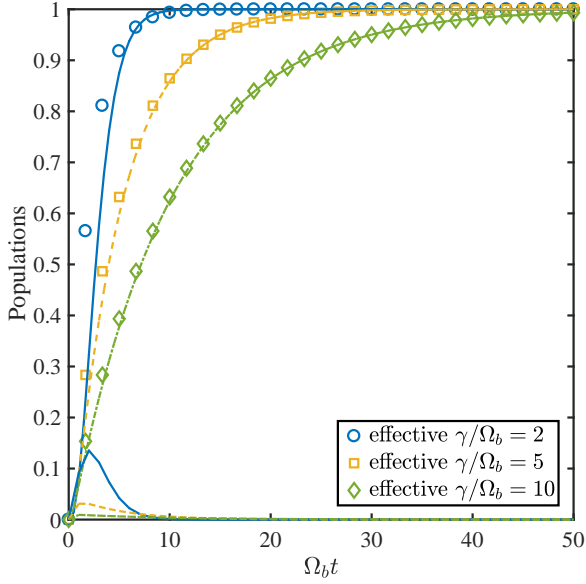} }
\caption{\label{spon1} Dynamic evolution of a single ion from the initial state $|r\rangle$ versus different spontaneous emission rates of state $|a\rangle$. The upper lines (solid, dashed, and dash-dotted) indicate the populations of the ground states, while the lower lines show the population of the short-lived state $|a\rangle$. The markers of circle, square, and diamond are the results derived from the effective master equation of Eq.~(\ref{spon}). }
\end{figure}
According to Eqs.~(\ref{eff2}) and (\ref{eff6}), a single-ion master equation describing the effective spontaneous decay of metastable state can be derived as
\begin{equation}
\dot{\hat\rho}_{\rm sin} = \frac{\gamma_{\rm eff}}{2}\hat {\cal L}[|g\rangle\langle r|]\hat\rho_{\rm sin}+\frac{\gamma_{\rm eff}}{2}\hat {\cal L}[|e\rangle\langle r|]\hat\rho_{\rm sin}. \label{spon}
\end{equation}

Fig.~\ref{spon1} displays the time evolution of a single ion with the initial state $|r\rangle$ versus different spontaneous emission rates of state $|a\rangle$. The upper solid, dashed, and dash-dotted lines indicate the populations of the ground states $\langle g|{\hat\rho}_{\rm sin}|g\rangle+\langle e|{\hat\rho}_{\rm sin}|e\rangle$, corresponding to the decay rates $\gamma/\Omega_b=2$, $\gamma/\Omega_b=5$, and $\gamma/\Omega_b=10$, respectively. Meanwhile, the results obtained from the effective master equation are demonstrated by the markers of circle, square, and diamond. Now we see that Eq.~(\ref{spon}) is sufficient to describe the dynamic evolution of Eq.~(\ref{iMastEqns}) as long as $\gamma/\Omega_b\geq5$, since the population of state $|a\rangle$ can be eliminated adiabatically.

Combination of the results from Eq.~(\ref{Mast}) and  Eq.~(\ref{spon}), we are able to achieve the dissipative dynamics of two effective three-level ions with a controllable spontaneous emission rate $\gamma_{\rm eff}/2$, i.e.
\begin{equation}
\dot{\hat\rho}_t = i[\hat\rho_t,{\hat H_{\rm eff}}]+\frac{\gamma_{\rm eff}}{2}\sum_{j=1}^2 \{\hat {\cal L}[|g_j\rangle\langle r_j|]\hat\rho_t+\hat {\cal L}[|e_j\rangle\langle r_j|]\hat\rho_t\}. \label{MastEqn}
\end{equation}

\begin{figure}
\scalebox{0.42}{\includegraphics{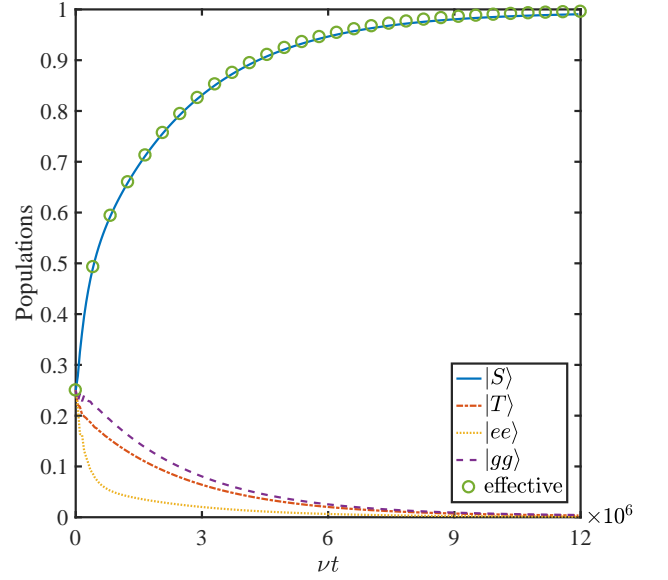} }
\caption{\label{p2}Populations for different qubit states are plotted as functions of time. The ions are initialized in a fully mixed state $\rho_0=\Sigma_{k,l=g,e}|kl\rangle\langle kl|/4$ together with a mixed occupation of the vibrational modes $\prod_{p=1}^2(|0_p\rangle\langle 0_p|+|1_p\rangle\langle 1_p|)/2$.  The other relevant parameters are set as $\nu/2\pi=2$~MHz, $\eta=0.1$, $\Omega_a/2\pi=200$~kHz, $\gamma_{\rm eff}/2\pi=0.2$~kHz, and $\Omega_{mw}=-2\lambda$.}
\end{figure}

To illustrate the reasonableness of the above derivation, we use the definition of the population $P=\langle X|{\rm Tr}_{n_1,n_2}[\hat\rho_t]|X\rangle$ and plot its dependence on time in Fig.~\ref{p2}, where $X=S, T, ee, gg$ labels the internal state of two ions and ${\rm Tr}_{n_1,n_2}$ denotes the partial trace over the unobserved
vibrational degrees of freedoms. The four different lines (solid, dashed, dotted, and dash-dotted) are simulated with {a full Hamiltonian of Eq.~(\ref{full}).
Initially, the ions are assumed to be in a fully mixed state $\rho_0=\Sigma_{k,l=g,e}|kl\rangle\langle kl|/4$ and the vibrational modes are in a mixed state {$\prod_{p=1}^2(|0_p\rangle\langle 0_p|+|1_p\rangle\langle 1_p|)/2$}, which are then cut off by $n_p=2$ for the sake of simplicity. The empty circle represents the population of the target Bell state obtained from the effective master equation of Eq.~(\ref{MastEqn}), which is in an excellent agreement
with the solid line under the given parameters $\nu/2\pi=2$~MHz, $\eta=0.1$, $\Omega_a/2\pi=200$~kHz, $\gamma_{\rm eff}/2\pi=0.2$~kHz, and $\Omega_{mw}=-2\lambda$. Note that for different trapped-ion systems, the frequencies of the center-of-mass mode may be different, but we are still allowed to prepare a high-fidelity antisymmetric Bell state in a relative short time, since the Rabi frequencies of laser pulses are adjustable.

\section{Effect of temperature}
 So far, we have discussed how to prepare a maximally entangled state through engineering the spontaneous emission rates of trapped ions from an arbitrary initial state, regardless of details of the vibrational modes. This feature will no doubt simplify the experimental operations compared to previous schemes where the quantum bus has to be initialized in the vacuum state \cite{PhysRevLett.106.090502,shen2011steady,PhysRevA.96.062315,Li:18}. From the viewpoint of steady state, this phonon-independent proposal may also favor a high quality of entanglement for a finite-temperature bath. The steady-state solution of the whole system can be found by solving the following master equation,
\begin{eqnarray}
0&=& i[\hat\rho_\infty,{\hat H_I}]+\frac{\gamma_{\rm eff}}{2}\sum_{j=1}^2 \{\hat {\cal L}[|g_j\rangle\langle r_j|]\hat\rho_\infty+\hat {\cal L}[|e_j\rangle\langle r_j|]\hat\rho_\infty\}\nonumber\\
&&+\sum_{p=1}^2\{\kappa_p(\bar{n}^p_{\rm th}+1)\hat {\cal L}[\hat a_p]\hat\rho_\infty+\kappa_p(\bar{n}^p_{\rm th})\hat {\cal L}[\hat a_p^{\dag}]\hat\rho_\infty\}, \label{MastEqnful}
\end{eqnarray}
where $\kappa_p$ characterizes the decay rate of the $p$th vibrational mode and $\bar{n}^p_{\rm th}$ denotes the mean vibrational number of the reservoir in the thermal equilibrium. Here we apply the Clauser-Horne-Shimony-Holt (CHSH) correlation $
S={\rm Tr}\big[({\cal \hat O}_{\rm CHSH}){\rm Tr}_n[\hat\rho_{\infty}]\big],
$ to assess the performance of the current scheme, where the operator ${\cal \hat O}_{\rm CHSH}$ is defined as \cite{PhysRevLett.23.880}
\begin{eqnarray}
{\cal \hat O}_{\rm CHSH}&=&\hat\sigma_{y,1}\otimes\frac{-\hat\sigma_{y,2}-\hat\sigma_{x,2}}{\sqrt{2}}
+\hat\sigma_{x,1}\otimes\frac{-\hat\sigma_{y,2}-\hat\sigma_{x,2}}{\sqrt{2}}\nonumber\\
&&+\hat\sigma_{x,1}\otimes\frac{\hat\sigma_{y,2}-\hat\sigma_{x,2}}{\sqrt{2}}
-\hat\sigma_{y,1}\otimes\frac{\hat\sigma_{y,2}-\hat\sigma_{x,2}}{\sqrt{2}}.\label{chsh}
\end{eqnarray}
Concerning our model, the above pauli matrices are written in the basis of $\{|g\rangle=(1,0)^T, |e\rangle=(0,1)^T\}$.
{The Bell-CHSH inequality is a useful theoretical tool for demonstrating quantum nonlocality. Quantum mechanics violates Bell inequality when the CHSH correlation rises above 2,
and a singlet state of two spin$-1/2$ particles maximally  violates the CHSH correlation which can reach the Tsirelson's bound of $S=2\sqrt{2}$.}

\begin{figure}
\scalebox{0.337}{\includegraphics{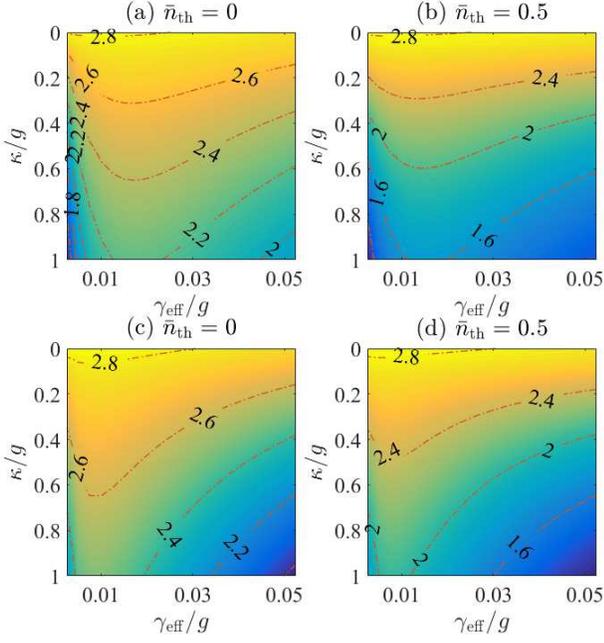} }
\caption{\label{p3} Contour plot of the steady-state CHSH correlation versus ionic spontaneous emission $\gamma_{\rm eff}/g$ and phonon decay $\kappa/g$ at different temperatures of bath. The frequency of the center-of-mass mode is $\nu/2\pi=2$~MHz  for (a) $\bar{n}_{\rm th}=0$ and (b) $\bar{n}_{\rm th}=0.5$, and $\nu/2\pi=4$~MHz for (c) $\bar{n}_{\rm th}=0$ and (d) $\bar{n}_{\rm th}=0.5$. Other parameters are $\eta=0.1$, $\Omega_a/2\pi=200$~kHz, and $\Omega_{mw}=-2\lambda$.}
\end{figure}
In Fig.~\ref{p3}, we show the contour plot of the steady-state CHSH correlation versus ionic spontaneous emission $\gamma_{\rm eff}/g$ ($g=\eta\Omega_a/2$) and phonon decay $\kappa/g$ for two frequencies of the center-of-mass mode at different temperatures of bath, where $\kappa_a=10\kappa_b=\kappa$ and $\bar{n}^p_{\rm th}=\bar{n}_{\rm th}$ are assumed since only the center-of-mass
mode will be heated up by spatially coherent fields \cite{PhysRevLett.81.317}. We choose
 $\nu/2\pi=2$~MHz  in (a) $\bar{n}_{\rm th}=0$ and (b) $\bar{n}_{\rm th}=0.5$, and $\nu/2\pi=4$~MHz in (c) $\bar{n}_{\rm th}=0$ and (d) $\bar{n}_{\rm th}=0.5$, and other parameters are $\eta=0.1$, $\Omega_a/2\pi=200$~kHz, and $\Omega_{mw}=-2\lambda$. A direct comparison shows that the temperature of the reservoir has a detrimental effect in entanglement, but the ions still maintain the quantum correlation for a wide range of the decoherence parameters. Furthermore, the effect of phonon decay can be reduced by choosing a larger frequency of center-of-mass mode, because the decoupling between ions and the vibrational modes becomes more efficient in this condition due to the requirement of Eq.~(\ref{effective}).

\section{imperfect alignment of the ions}
The previous discussion is on the basis of a perfect alignment of the ions. Nevertheless, the effect of misalignment is unavoidable in experiments. In this part, we will analyze this effect on our proposal and then put forward the way to overcome it. Supposing the ions are positioned away from the node of laser field by a small deviation ($\phi\ll1$), and then the laser-ion interaction Hamiltonian of Eq.~(\ref{xxx}) is given by \cite{PhysRevLett.90.217901}
\begin{eqnarray}\label{be}
{\hat H_i}&=&\Omega_a\sum_{j=1}^2\hat s_{j,x}\bigg[\sum_{p=1}^2\eta_{jp}({\hat a_p^{\dag}}e^{i\nu_p t}+{\hat a_p}e^{-i\nu_p t})\cos\phi\nonumber\\
&&+\sin\phi\bigg],
\end{eqnarray}
where $\hat s_{j,x}={(|e_j\rangle\langle r_j|+|r_j\rangle\langle e_j|)}/{2}$. The complete evolution operator induced by this Hamiltonian takes the form of
\begin{eqnarray}\label{unitary}
{\hat U}(t)&=&\prod_{j=1}^2e^{-i\Omega_a\sin\phi{\hat s_{j,x}}t}\prod_{p=1}^2e^{-iA_p(t){\hat J^2_{p,x}}}e^{-iB_p(t){\hat J_{p,x}}{\hat a_p}}\nonumber\\
&&\times e^{-iC_p(t){\hat J_{p,x}}{\hat a^{\dag}_p}}.
\end{eqnarray}
where $
{\hat J_{p,x}}=\sum_{j=1}^2\eta_{jp}{\hat s_{j,x}}/\eta
$ and

\begin{eqnarray}\label{2}
A_p(t)=\eta^2\Omega_a^2\cos^2\phi\bigg[-\frac{1}{\nu_p}t+\frac{1}{i\nu_p^2}(e^{i\nu_pt}-1)\bigg],
\end{eqnarray}
\begin{eqnarray}\label{2}
B_p(t)=\frac{\eta\Omega_a}{-i\nu_p}\cos\phi(e^{-i\nu_pt}-1),
\end{eqnarray}
\begin{eqnarray}\label{2}
C_p(t)=\frac{\eta\Omega_a}{i\nu_p}\cos\phi(e^{i\nu_pt}-1).
\end{eqnarray}
\begin{figure}
\scalebox{0.42}{\includegraphics{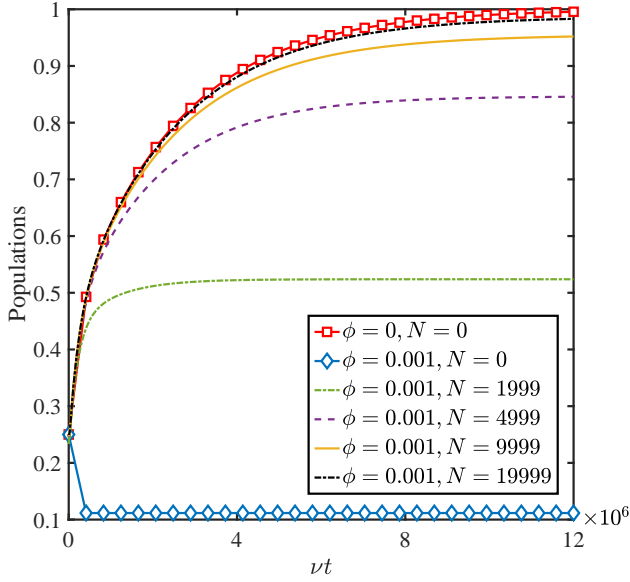} }
\caption{\label{swap} Populations of the target Bell state with different switching times for the phase of the laser field. The initial state and relevant parameters are the same as in Fig.~\ref{p2}.}
\end{figure}

Under the condition $\eta\Omega_a/2\ll\nu_p$, $B_p(t)$, $C_p(t)$ and the seconde part of $A_p(t)$ becomes negligible. Thus the evolution operator of Eq.~(\ref{unitary}) can be considered as governed by an effective Hamiltonian of
\begin{eqnarray}\label{phi}
{\hat H_{i}}&\approx&-\sum_{p=1}^2\bigg\{\frac{{\eta_{1p}\eta_{2p}\Omega_a^2}}{2{\nu_p}}\prod_{j=1}^2(|e_j\rangle\langle r_j|+|r_j\rangle\langle e_j|)\nonumber\\
&&+\sum_{j=1}^2\bigg[\frac{{\eta_{jp}^2\Omega_a^2}}{4{\nu_p}}(|e_j\rangle\langle e_j|+|g_j\rangle\langle g_j|)\bigg]\bigg\}\cos^2\phi\nonumber\\
&&+\Omega_a\sum_{j=1}^2\hat s_{j,x}\sin\phi.
\end{eqnarray}
For $\phi=0$, the result of Eq.~(\ref{effective}) is recovered. Now we replace the effective Hamiltonian $\hat H_{\rm eff}$ of Eq.~(\ref{MastEqn})
with Eq.~(\ref{phi}) and choose the same initial state and relevant parameters as in Fig.~\ref{p2} to plot the population of the target Bell state in Fig.~\ref{swap}. We see that
even a very small error ($\phi=0.001$) in the positioning of the ions introduces a very large effect (the population reduces to $11.15\%$), as shown by the empty diamonds. Fortunately, the effect of misalignment can be suppressed in an experimentally simple way. Back to Eq.~(\ref{xxx}), if the phase of the laser field is chosen as $\phi^{'}=\phi+\pi$, the laser-ion interaction Hamiltonian becomes
\begin{eqnarray}\label{2}
{\hat H_i^{'}}&=&-\Omega_a\sum_{j=1}^2\hat s_{j,x}\bigg[\sum_{p=1}^2\eta_{jp}({\hat a_p^{\dag}}e^{i\nu_p t}+{\hat a_p}e^{-i\nu_p t})\cos\phi\nonumber\\
&&+\sin\phi\bigg].
\end{eqnarray}
This operation corresponds to exchanging the signs
of $\Omega_a$. Similar to the process from Eq.~(\ref{be}) to Eq.~(\ref{phi}), we have the effective interaction Hamiltonian in this case as
\begin{eqnarray}\label{phii}
{\hat H_{i}^{'}}&\approx&-\sum_{p=1}^2\bigg\{\frac{{\eta_{1p}\eta_{2p}\Omega_a^2}}{2{\nu_p}}\prod_{j=1}^2(|e_j\rangle\langle r_j|+|r_j\rangle\langle e_j|)\nonumber\\
&&+\sum_{j=1}^2\bigg[\frac{{\eta_{jp}^2\Omega_a^2}}{4{\nu_p}}(|e_j\rangle\langle e_j|+|g_j\rangle\langle g_j|)\bigg]\bigg\}\cos^2\phi\nonumber\\
&&-\Omega_a\sum_{j=1}^2\hat s_{j,x}\sin\phi.
\end{eqnarray}

A comparison of  Eq.~(\ref{phi}) and Eq.~(\ref{phii}) indicates that the extra term  $\exp(-i\Omega_a\tau\sum_{j=1}^2\hat s_{j,x}\sin\phi)$
originated from illuminating the ions with a laser
of phase $\phi$ for a time $\tau$ can be canceled out by the term $\exp(i\Omega_a\tau\sum_{j=1}^2\hat s_{j,x}\sin\phi)$ arising from subsequently illuminating the ions with a laser
of phase $\phi^{'}$ for another time $\tau$. This ``photon-echo"-like procedure has been successfully applied in the unitary-based schemes of trapped ions to cancel the motion-dependent term in Hamiltonian \cite{PhysRevLett.87.127901,PhysRevLett.90.217901}. However, for the current dissipation-based scheme, a few switching times of the laser phase cannot play a significant role, since the dynamics of the whole system is irreversible.

Using the principle of quantum Zeno effect for reference, we divide the convergence time for the perfect alignment case into $(N+1)$ equal pieces and plot the populations of the target Bell state with different switching times $N$ in Fig.~\ref{swap} in the presence of $\phi=0.001$. It displays that the population of the Bell state has been significantly improved with $N=1999$, and this value can further reach $98.31\%$ as $N=19999$. Therefore, it is feasible in principle to eliminate the influence of misalignment of ions by frequently switching the phase of the laser field.
\begin{figure}
\scalebox{0.315}{\includegraphics{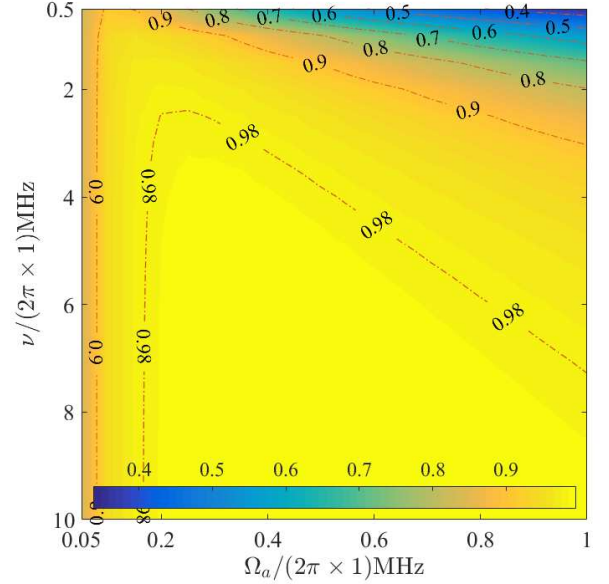} }
\caption{\label{flu} Population of the target Bell state in the steady state versus different frequencies of center-of-mass mode and Rabi frequencies of laser field. Other parameters: $\eta=0.1$, $\gamma_{\rm eff}/2\pi=0.2$~kHz, $\kappa_1/2\pi=1$~kHz, $\kappa_2/2\pi=0.1$~kHz, $\bar{n}_{\rm th}=0$, and $\Omega_{mw}=-2\lambda$.}
\end{figure}

\section{other noise and experimental feasibility}
The fluctuations of control fields and center-of-mass mode are also inevitable factors in experiments of trapped ions. Especially for the scheme requiring the particular values of Rabi frequency of the laser field \cite{PhysRevLett.87.127901}, where a small fluctuation of Rabi frequency can cause a deviation from the resonance conditions. On the contrary, the current proposal only demands $\eta_{jp}\Omega_a/2\ll\nu_p$, and there is no special requirement for $\Omega_a$. In Fig.~\ref{flu}, we investigate the population of the target Bell state in the steady state as a function of $\Omega_a$ and $\nu$, which shows that our scheme is robust against the fluctuations of these two parameters since the population exceeds $98\%$ over a wide area.

The fluctuations in control fields that couple to quantum systems globally can lead to the decoherence of collective dephasing \cite{PhysRevLett.106.130506,schindler2013quantum}. To study this effect on the performance of the entanglement preparation, we incorporate the collective dephasing operator into Eq.~(\ref{MastEqnful}) and have the steady-state master equation as

\begin{figure}
\scalebox{0.42}{\includegraphics{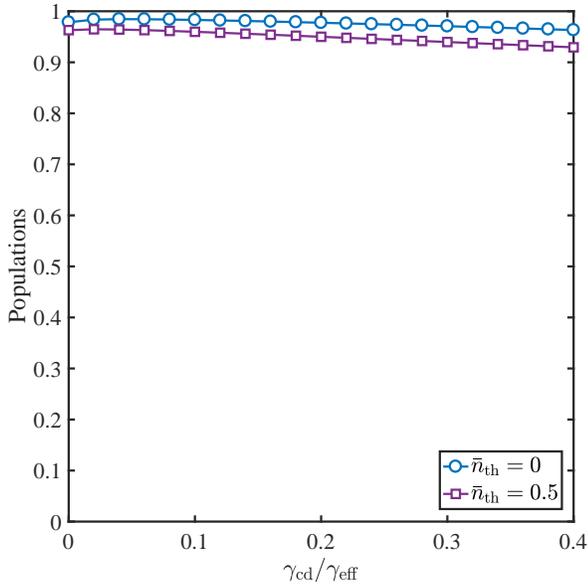} }
\caption{\label{dep} Population of the target Bell state in the steady state versus different strengths of collective dephasing rate. Other parameters: $\nu/2\pi=2$~MHz, $\Omega_a/2\pi=200$~kHz, $\kappa_1/2\pi=1$~kHz, $\kappa_2/2\pi=0.1$~kHz, $\gamma_{\rm eff}/2\pi=0.2$~kHz, and $\Omega_{mw}=-2\lambda$.}
\end{figure}
\begin{eqnarray}
0=\hat {\cal L}_{\rm all}\hat\rho_\infty +\gamma_{\rm cd}\hat {\cal L}\bigg[\sum_{j=1}^2 (|g_j\rangle\langle g_j|-|r_j\rangle\langle r_j|)\bigg]\hat\rho_\infty, \label{MastEqnful1}
\end{eqnarray}
where $\hat {\cal L}_{\rm all}\hat\rho_\infty$ has included all terms in Eq.~(\ref{MastEqnful}), and $\gamma_{\rm cd}$ means the collective dephasing rate. The steady-state population is plotted as a function of $\gamma_{\rm cd}/\gamma_{\rm eff}$ in Fig.~\ref{dep} at different temperatures of reservoir, which remains to be higher than $90\%$ as $\gamma_{\rm cd}=0.4\gamma_{\rm eff}$ under the given parameters.

The experimental setup for the present scheme may employ $^{40}{\rm C_a^{+}}$ ions held in a linear Paul trap, where various motional frequencies from $\nu/2\pi=2$~MHz to $\nu/2\pi=4$~MHz are achieved \cite{PhysRevLett.83.4713,haffner2008quantum}. Two Zeeman substates of the
$4S_{1/2}$ ground states are encoded as $4S_{1/2}(m =-1/2) = |g\rangle$ and $4S_{1/2}(m =+1/2) = |e\rangle$, the $3D_{5/2}$ metastable state with the natural lifetime of 1.1$s$ corresponds to $3D_{5/2}= |r\rangle$, and the $4P_{3/2}$ level of lifetime $10^{-8}s$ serves as the excited short-lived state $4P_{3/2}= |a \rangle$.
The precision measurements predict that the branching ratios of the $4P_{3/2}$ level decay of a
single $^{40}{\rm Ca^+}$ ion into the $4S_{1/2}$, $3D_{5/2}$ and $3D_{3/2}$ levels are 0.9347(3), 0.0587(2)
and 0.00661(4), respectively \cite{Gerritsma2008}. And for convenience, we further simplify the fractions as $p_{S_{1/2}}\approx0.94$ and $p_{D_{5/2}}\approx0.06$. By considering the above experimental parameters
 in the company of $\nu/2\pi=4~$MHz, $\Omega_a/2\pi=200$~kHz, $\Omega_b/2\pi=40$~kHz, $\eta=0.1$, $\Omega_{mw}=-2\lambda$, and $\kappa_1=10\kappa_2=2\times10^{-4}\nu$, we accomplish the steady-state entanglement with populations of $98.90\%$ and $98.69\%$ corresponding to the average phonon number in the reservoir $\bar{n}_{\rm th}=0$ and $\bar{n}_{\rm th}=0.5$, respectively.

\section{summary}
In summary, we have found a new way to dissipatively prepare a maximally entangled
steady state in the trapped-ion system. This scenario makes use of the engineered spontaneous emission of internal ionic state as a resource while keeps decoupled from external vibrational freedoms. The effectiveness is confirmed from both sides of temporal evolution and steady state using Markovian master equation. We hope that this work may open new venues for the
experimental realization of steady entanglement in the near future.

\acknowledgements
The author thank the anonymous reviewer for constructive
comments that helped in improving the quality of this paper. This work is supported by National Natural Science Foundation of China (NSFC) under Grants No. 11774047.

\bibliography{new_ion_dissipation}

\end{document}